\documentclass[12pt]{article}
\usepackage{amssymb}
\usepackage{amsmath}

\newcommand{\be}{\begin{eqnarray}}
\newcommand{\ee}{\end{eqnarray}}
\newcommand{\non}{\nonumber}

\newcommand{\tr}{\mathop{\rm tr}\nolimits}

\begin{document}

\begin{titlepage}
\strut\hfill UMTG--276
\vspace{.5in}
\begin{center}

\LARGE Inhomogeneous $T$-$Q$ equation for the open XXX chain\\
with general boundary terms: completeness and arbitrary spin\\
\vspace{1in}
\large Rafael I. Nepomechie \footnote{nepomechie@physics.miami.edu}\\[0.8in]
\large Physics Department, P.O. Box 248046, University of Miami\\[0.2in]  
\large Coral Gables, FL 33124 USA\\

\end{center}

\vspace{.5in}

\begin{abstract}
An inhomogeneous $T$-$Q$ equation has recently been proposed by Cao,
Yang, Shi and Wang for the open spin-1/2 XXX chain with general
(nondiagonal) boundary terms.  We argue that a simplified version of
this equation describes all the eigenvalues of the transfer matrix of
this model.  We also propose a generating function for the
inhomogeneous $T$-$Q$ equations of arbitrary spin.
\end{abstract}

\end{titlepage}

\setcounter{footnote}{0}

\section{Introduction}\label{sec:intro}

The open spin-1/2 XXZ quantum spin chain with general (nondiagonal)
boundary terms has long been known to be integrable
\cite{Sklyanin:1988yz, Ghoshal:1993tm, deVega:1993xi}.  Nevertheless,
the Bethe ansatz solution of this model has remained elusive. \footnote{Solutions 
have been known for various special cases, such as
diagonal boundary terms \cite{ Sklyanin:1988yz, Gaudin:1971zza, Gaudin:1983,
Alcaraz:1987uk} and nondiagonal boundary terms whose
boundary parameters obey certain constraints \cite{Cao:2003,
Nepomechie:2003vv, Nepomechie:2003ez, Yang:2005ce}.} One of the 
difficulties is that a reference state (simple eigenstate of the 
transfer matrix) is not available. 

Another (perhaps related) difficulty concerns the so-called Baxter
$T$-$Q$ equation.  For the periodic spin-1/2 XXX chain, given an
eigenvalue $T$ of the transfer matrix (which is a polynomial function
of the spectral parameter), the $T$-$Q$ equation is a homogeneous
linear second-order finite-difference equation for $Q$, which is also
a polynomial function of the spectral parameter.  However, for the
corresponding open chain with general boundary terms, the asymptotic
behavior of $T$ for large values of the spectral parameter is not
compatible with both (i) a conventional (homogeneous) $T$-$Q$ equation
and (ii) $Q$-functions that are polynomial functions of the spectral
parameter (see e.g. \cite{Frahm:2010}).

An important advance was recently made by Cao, Yang, Shi and Wang
(CYSW) \cite{Cao:2013qxa, Cao:2013bca} (see also \cite{Cao:2013nza,
Cao:2013gug}), who proposed to use instead an {\it inhomogeneous}
$T$-$Q$ equation.  Although the inhomogeneous term in the $T$-$Q$
equation gives rise to an unconventional term in the Bethe equations,
it overcomes the difficulty of reconciling the asymptotic behavior of
$T$ with $Q$'s that are polynomial functions of the spectral
parameter.

Several interesting questions about this approach remain to be
addressed, and in this note we focus on two of them: the completeness
of the proposed solution, and the generalization to higher spin.  For
simplicity, we restrict to the isotropic (XXX) chain.  After briefly
reviewing in Section \ref{sec:review} the solution proposed by CYSW
\cite{Cao:2013qxa}, we argue in Section \ref{sec:completeness} that a
simplified version of their solution is already complete.  Hence,
every eigenvalue of the transfer matrix can be characterized by a set
of Bethe roots.  In Section \ref{sec:higherspin}, we formulate the
inhomogeneous $T$-$Q$ equation for an auxiliary space of arbitrary
spin.  We briefly summarize our conclusions and mention some
outstanding problems in Section \ref{sec:conclusion}.

\section{The CYSW solution}\label{sec:review}

Following \cite{Cao:2013qxa}, we consider an open spin-1/2 XXX chain 
of length $N$, with the Hamiltonian
\be
H =  \sum_{n=1}^{N-1}  
\vec \sigma_{n} \cdot \vec \sigma_{n+1} + \frac{1}{p}\sigma^{z}_{N} 
+ \frac{1}{q}\left(\sigma^{z}_{1} +\xi \sigma^{x}_{1}\right)\,,
\label{Hamiltonian}
\ee 
where $p, q, \xi$ are arbitrary boundary parameters.
The corresponding transfer matrix is given by \cite{Sklyanin:1988yz}
\be
t(u) = \tr_{0} K^{+}_{0}(u)\, T_{0}(u)\, K^{-}_{0}(u)\, \hat T_{0}(u) \,,
\label{transfer}
\ee
where the monodromy matrices are given by \footnote{For simplicity, we set to zero the 
inhomogeneity parameters $\theta_{1}, \ldots \theta_{N}$ appearing in 
\cite{Cao:2013qxa}.}
\be
T_{0}(u) = R_{01}(u) \cdots R_{0N}(u)\,, \qquad \hat T_{0}(u) =
R_{0N}(u) \cdots R_{01}(u)
\,.
\ee
The $R$-matrix is given by
\be
R(u) = u + {\cal P} \,,
\ee 
where ${\cal P}$ is the permutation matrix, 
and the $K$-matrices are given by \cite{Ghoshal:1993tm, deVega:1993xi}
\be
K^{-}(u)=\left(\begin{array}{cc}
    p+u & 0 \\
    0 & p-u 
\end{array}    \right)\,, \qquad 
K^{+}(u)=\left(\begin{array}{cc}
    q+u+1 & \xi(u+1) \\
    \xi(u+1) & q-u-1 
\end{array}    \right) \,.
\ee 
The transfer matrix has the commutativity property $\left[t(u)\,, 
t(v)\right]=0$, and is therefore the generating function of a family 
of commuting operators, among which is the Hamiltonian (\ref{Hamiltonian}).

CYSW proposed the following inhomogeneous $T$-$Q$ equation:
\be
\Lambda^{(\pm)}(u)\, Q^{(\pm)}(u)\, Q^{(\pm)}_{1}(u)\, Q^{(\pm)}_{2}(u)
&=& \bar{a}^{(\pm)}(u)\, Q^{(\pm)}(u-1)\, Q^{(\pm)}_{1}(u-1)\, 
Q^{(\pm)}_{1}(u) \non \\
& & + \bar{d}^{(\pm)}(u)\, Q^{(\pm)}(u+1)\, Q^{(\pm)}_{2}(u+1)\, Q^{(\pm)}_{2}(u) \non \\
& & + 2(1-\sqrt{1+\xi^{2}})\left( u (u+1)\right)^{2N+1} \,, 
\label{TQ1}
\ee
where 
\be
\bar{a}^{(\pm)}(u) = \frac{2u+2}{2u+1}(u\pm p)(\sqrt{1+\xi^{2}} u \pm 
q)(u+1)^{2N}\,, \qquad \bar{d}^{(\pm)}(u) = \bar{a}^{(\pm)}(-u-1) \,,
\label{ad}
\ee  
and $\Lambda^{(\pm)}(u)$ are the eigenvalues of the transfer matrix 
(\ref{transfer}). Moreover, the $Q$'s are polynomial functions of $u$ given by
\be
Q^{(\pm)}(u) &=& 
\prod_{j=1}^{N-2M}(u-\lambda_{j}^{(\pm)})(u+\lambda_{j}^{(\pm)}+1) 
\,, \non \\
Q^{(\pm)}_{1}(u) &=& 
\prod_{j=1}^{M}(u-\mu_{j}^{(\pm)})(u+\nu_{j}^{(\pm)}+1) 
\,, \non \\
Q^{(\pm)}_{2}(u) &=& 
\prod_{j=1}^{M}(u-\nu_{j}^{(\pm)})(u+\mu_{j}^{(\pm)}+1) \,, 
\label{Qs}
\ee
where $M=0, \ldots, [\frac{N}{2}]$.
The key feature is the final (inhomogeneous) term in (\ref{TQ1}), 
which is absent if the Hamiltonian (\ref{Hamiltonian}) has only diagonal boundary terms 
($\xi =0$). As usual, the Bethe equations follow from the $T$-$Q$ 
equation and the requirement 
that the eigenvalues of the transfer matrix cannot have poles.
The eigenvalues of the Hamiltonian are given in terms of 
the Bethe roots (zeros of $Q$-functions) by
\be
E^{(\pm)} = 
2\sum_{j=1}^{N-2M}\frac{1}{\lambda_{j}^{(\pm)}(\lambda_{j}^{(\pm)}+1)}
+2\sum_{j=1}^{M}\left(\frac{1}{\nu_{j}^{(\pm)}}-\frac{1}{\mu_{j}^{(\pm)}+1}\right)+ c\,.
\label{energy1}
\ee 

\section{Completeness}\label{sec:completeness}

An important question is whether this solution is complete; i.e., 
whether every eigenvalue of the transfer 
matrix (\ref{transfer}) can be expressed in the form (\ref{TQ1}). 
CYSW  
conjectured \cite{Cao:2013qxa} that completeness is achieved by taking 
$M=\frac{N}{2}$ for $N$ even ($M=\frac{N+1}{2}$ for $N$ odd) in 
(\ref{Qs}), and by also
considering both signs $(\pm)$ in Eqs. (\ref{TQ1})-(\ref{energy1}), as in \cite{Nepomechie:2003ez}.

We conjecture that the situation is considerably simpler: it is enough to 
take $M=0$ in (\ref{Qs}), and consider just one sign (say, $+$) in 
Eqs. (\ref{TQ1})-(\ref{energy1}). In other words, we claim that 
completeness can be achieved (for any value of $N$) with the following linear $T$-$Q$ equation:
\be
\Lambda(u)\, Q(u) 
= \bar{a}(u)\, Q(u-1) + \bar{d}(u)\, Q(u+1)
 + 2(1-\sqrt{1+\xi^{2}})\left( u (u+1)\right)^{2N+1} \,, 
\label{TQ2}
\ee
where 
\be
\bar{a}(u) = \frac{2u+2}{2u+1}(u+ p)(\sqrt{1+\xi^{2}} u +
q)(u+1)^{2N} \,, \qquad \bar{d}(u) = \bar{a}(-u-1) \,,
\label{ad}
\ee  
and
\be
Q(u) = 
\prod_{j=1}^{N}(u-\lambda_{j})(u+\lambda_{j}+1)  \,.
\label{Q}
\ee
The eigenvalues of the Hamiltonian (\ref{Hamiltonian}) are then given by
\be
E = 
2\sum_{j=1}^{N}\frac{1}{\lambda_{j}(\lambda_{j}+1)} +c\,,
\label{energy2}
\ee 
where $c=N-1+\frac{1}{p}+\frac{\sqrt{1+\xi^{2}}}{q}$.

This conjecture is supported by numerical evidence for chains of small
length.  Indeed, for generic values of the boundary parameters and $N=
2, \ldots, 8$, we have determined (following the method described
in \cite{Nepomechie:2003ez}) the set of Bethe roots $\{ \lambda_{1},
\ldots, \lambda_{N}\}$ that characterize each of the $2^{N}$
eigenvalues via Eqs.  (\ref{TQ2})-(\ref{Q}).  As a check, we have
verified that the energies computed using (\ref{energy2}) coincide
with those obtained by direct diagonalization of the Hamiltonian
(\ref{Hamiltonian}).  Sample results for $N=3$ and $N=4$ are presented
in Tables \ref{table:N=3} and \ref{table:N=4}, respectively.  

We have also numerically verified completeness using instead the $T$-$Q$
equation (\ref{TQ1}) with $M=1$ and again with just one sign (say, 
$+$) for $N=2, 3$.  However, that equation
is nonlinear in the $Q$'s, and is therefore significantly more
complicated than (\ref{TQ2}), which is linear in $Q$.

\section{Higher spin}\label{sec:higherspin}

The transfer matrix (\ref{transfer}) is constructed by tracing over a
spin-1/2 (i.e., two-dimensional) auxiliary space.  Using the fusion
procedure \cite{Kulish:1981gi}, open-chain transfer matrices
$T_{1,s}(u)$ corresponding to auxiliary spaces of arbitrary spin 
$s/2$  ($s = 1, 2, \ldots$) can be constructed \cite{Mezincescu:1990fc, Mezincescu:1991ke,
Zhou:1995zy}.  These transfer matrices (and therefore their 
eigenvalues) obey the Hirota equation (see e.g.  \cite{Gromov:2009tv} 
and references therein) \footnote{We follow the notation in \cite{Gromov:2009tv} such that, 
for any function $f(u)$,  
$f^{\pm} = f(u\pm \frac{i}{2})$ and $f^{[\pm n]} = f(u\pm \frac{i n}{2})$.}
\be
T_{1, s}^{+}\, T_{1,s}^{-} = T_{2,s} + T_{1,s+1}\, T_{1,s-1} \,, 
\qquad s = 1, 2, \ldots \,,
\label{Hirota}
\ee 
where $T_{2,s}$ satisfies 
\be
T_{2, s}^{+}\, T_{2,s}^{-} =   T_{2,s+1}\, T_{2,s-1} \,,
\ee 
and $T_{a,0} = 1$.

For the special case of diagonal boundary terms ($\xi=0$), the eigenvalues of 
$T_{1,s}$ (which by abuse of notation we henceforth denote by the same symbol)
satisfy simple homogeneous $T$-$Q$ relations, which can be compactly formulated in 
terms of the following generating functional:\footnote{The 
eigenvalues of the corresponding closed-chain transfer matrices 
have essentially the same structure. Hence, the generating function 
(\ref{W1}) has presumably already appeared in the literature, but 
we have not been able to find the reference.}
\be
{\cal W}_{diag} = (1 - {\cal D} B {\cal D})^{-1}\, (1 - {\cal D} A {\cal D})^{-1} = 
\sum_{s=0}^{\infty} {\cal D}^{s}\, T_{1,s} \,  {\cal D}^{s} \,,
\label{W1}
\ee
where 
\be
A = \bar{a}\, \frac{Q^{[-2]}}{Q} \,, \qquad B = \bar{d}\, \frac{Q^{[+2]}}{Q} \,,
\label{AB}
\ee
and ${\cal D} =e^{-\frac{i}{2}\partial_{u}}$ implying that ${\cal D} 
f = f^{-} {\cal D}$. For example, by expanding both sides of (\ref{W1}), 
one easily obtains 
\be
T_{1,0} &=& 1 \,, \\
T_{1,1} &=& A + B = \bar{a}\, \frac{Q^{[-2]}}{Q} + \bar{d}\, \frac{Q^{[+2]}}{Q} \,, \label{s1} \\
T_{1,2} &=& A^{+} A^{-} + A^{-} B^{+} + B^{+} B^{-} \non \\ 
&=& \bar{a}^{+}\bar{a}^{-} \frac{Q^{[-3]}}{Q^{+}} + 
\bar{a}^{-} \bar{d}^{+} \frac{Q^{[+3]}Q^{[-3]}}{Q^{+} Q^{-}} + 
\bar{d}^{+}\bar{d}^{-} \frac{Q^{[+3]}}{Q^{-}}
\,. \label{s2} 
\ee 
Eq. (\ref{s1}) is a rewriting of the fundamental (spin-1/2) $T$-$Q$ equation 
(\ref{TQ2}), while  (\ref{s2}) is the spin-1 $T$-$Q$ equation. These 
transfer matrix eigenvalues satisfy the Hirota equation 
(\ref{Hirota}) with
\be
T_{2,s} =\prod_{k=-\frac{s-1}{2}}^{\frac{s-1}{2}}A^{[2k+1]} 
B^{[-2k-1]} = \prod_{k=-\frac{s-1}{2}}^{\frac{s-1}{2}}\bar{a}^{[2k+1]} 
\bar{d}^{[-2k-1]} \,,
\label{T2s}
\ee 
which is independent of $Q$.

For the general case of nondiagonal boundary terms, we propose the 
following generating function for the inhomogeneous $T$-$Q$ equations:
\be
{\cal W} = \left[1 - {\cal D} (A+B+C) {\cal D} + {\cal D} A {\cal 
D}^{2} B {\cal D} \right]^{-1} = 
\sum_{s=0}^{\infty} {\cal D}^{s}\, T_{1,s} \,  {\cal D}^{s} \,,
\label{W2}
\ee
where $A$ and $B$ are again given by (\ref{AB}), and $C$ is given by
\be
C=\frac{\Delta}{Q} \,,
\ee 
where $\Delta$ is the inhomogeneous term in the $T$-$Q$ equation (\ref{TQ2})
\be
\Delta= 2(1-\sqrt{1+\xi^{2}})\left( u (u+1)\right)^{2N+1} \,.
\ee 

Our proposal (\ref{W2}) for the generating function passes several
nontrivial checks.  First, for the special case of diagonal boundary
terms ($C=0)$, it correctly reduces to the previous result (\ref{W1}):
\be
{\cal W}\Big\vert_{C=0} &=& \left[1 - {\cal D} (A+B) {\cal D} + {\cal D} A {\cal 
D}^{2} B {\cal D} \right]^{-1} \non \\
&=& \left[ (1 - {\cal D} A {\cal D}) (1 - {\cal D} B {\cal D})\right]^{-1} \non \\
&=& (1 - {\cal D} B {\cal D})^{-1}\, (1 - {\cal D} A {\cal D})^{-1} = 
{\cal W}_{diag} \,.
\ee

Moreover, by expanding both sides of (\ref{W2}), we obtain
\be
T_{1,0} &=& 1 \,, \\
T_{1,1} &=& A + B + C = \bar{a}\, \frac{Q^{[-2]}}{Q} + \bar{d}\, 
\frac{Q^{[+2]}}{Q} + \frac{\Delta}{Q}\,, \label{ss1} \\
T_{1,2} &=&  A^{+} A^{-} + A^{-} B^{+} + B^{+} B^{-} + C^{+}(A^{-} + B^{-})
+ C^{-}(A^{+} + B^{+}) +  C^{+} C^{-} \,, \label{ss2} \\
&\vdots & \non 
\ee 
Eq. (\ref{ss1}) is a rewriting of the $T$-$Q$ equation (\ref{TQ2}), 
and  (\ref{ss2}) is our result for the spin-1 inhomogeneous $T$-$Q$ equation. We 
have explicitly verified that this expression for $T_{1,2}$ (and 
similarly for $T_{1,s}$ up to $s=4$) satisfies the 
Hirota equation (\ref{Hirota}), where $T_{2,s}$ is again given by 
(\ref{T2s}). 

Notice that the inhomogeneous $T$-$Q$ equation for spin 
1 (\ref{ss2}) has 8 terms (once all the products are expanded), while its homogeneous 
counterpart (\ref{s2}) has only 3. We find that the number of terms $n_{s}$ in the 
inhomogeneous $T_{1,s}$ is given by the following generating function:
\be
w = (1-3x+x^{2})^{-1} = 1 + 3x + 8 x^{2} + 21 x^{3} + \ldots = 
\sum_{s=0}^{\infty} n_{s} x^{s} \,.
\label{character}
\ee
As yet another check of the generating function ${\cal W}$ (\ref{W2}), we observe that it reduces to 
$w$  (\ref{character}) in the 
character limit ${\cal D} A {\cal D} = {\cal D} B {\cal D} = {\cal D} 
C {\cal D} \equiv x$.

\section{Conclusion}\label{sec:conclusion}

We have argued that a simplified version of the CYSW $T$-$Q$ equation,
namely (\ref{TQ2}), describes all the eigenvalues of the transfer
matrix of the open spin-1/2 XXX chain with general boundary terms.  We
hope that it, together with the corresponding Bethe equations, can now
be used to investigate the thermodynamic limit of this model.  Due to
the presence of several boundary parameters, we expect that this model
has a rich phase structure.  This solution may have many other
applications, such as the brane-antibrane system in AdS/CFT
\cite{Bajnok:2013xxx}.

We have also proposed a generating function (\ref{W2}) for the
inhomogeneous $T$-$Q$ equations of arbitrary spin $s/2$, $s = 1, 2,
\ldots$ These equations differ significantly from their familiar
homogeneous counterparts, and may be useful in more formal
investigations of the model.

It should be possible to generalize this work to anisotropic spin
chains \cite{Cao:2013bca}, and to integrable spin chains with bulk
symmetry algebras of higher rank.

We have focused here only on the eigenvalues of the model's transfer matrix.
For a recent discussion of also the eigenvectors, see
\cite{Faldella:2013qha} and references therein.

\section*{Acknowledgments}
It is a pleasure to thank Zoltan Bajnok, Marcio Martins, and Dmytro 
Volin for valuable
discussions, and the I.E.U. at Ewha University for hospitality during 
the course of this work.
This work was supported in part by the National Science Foundation 
under Grants PHY-0854366 and PHY-1212337, and by a Cooper fellowship.

   
\providecommand{\href}[2]{#2}\begingroup\raggedright\endgroup


\vspace{1cm}

\begin{table}[htb] 
  \centering
  \begin{tabular}{|c|c|}\hline
    $E$ & Bethe roots $\lambda_{j}$ \\
    \hline
     -10.4854 & $-0.301706, -0.228269, 1.90659 $\\
      -6.3650 & $-0.202149, 0.0000179 \pm 0.0760986 i $\\
       -1.6983 & $-0.5 - 1.36473 i, -0.234301, 1.80106 $ \\
        -0.5138 & $-0.5 - 1.35297 i, -0.278630, 1.79670 $ \\
	  0.8170  & $-0.244206, 0.829712, 1.99163 $\\
	    1.2142  & $-0.257109, 0.816209, 1.99041 $\\
	      7.2463  & $1.79880, -0.064122 \pm 0.726059 i$ \\
    9.78493 & $-0.5 + 2.64170 i, 1.75921 \pm 1.91745 i$  \\
    \hline
   \end{tabular}
   \caption{The complete set of $2^{N}$ energy levels and Bethe roots for 
   $N=3$
   using Eqs.  (\ref{TQ2})-(\ref{Q}) and boundary parameter values $p=1/4\,, q=1/2\,, \xi=-\sqrt{3}$.}
   \label{table:N=3}
\end{table}
\begin{table}[htb] 
  \centering
  \begin{tabular}{|c|c|}\hline
    $E$ & Bethe roots $\lambda_{j}$ \\
    \hline
      -11.7918 & $-0.5 + 0.929239 i, -0.267373, -0.239061, 2.63465$\\
       -9.54275 & $-0.260299, -0.242542, 1.59304, 2.39063$\\
         -7.52344 & $-0.2479 \pm 0.0109844 i, 0.498355, 2.56045$ \\
	  -5.21618 & $-0.288548, -0.233564, 0.383076 \pm 0.239546 i$\\
	   -3.62486 & $-0.5 + 0.427847 i, -0.243628, 1.37433, 2.46576$\\
	     -3.21479 & $-0.5 + 0.426563 i, -0.25838, 1.36582, 2.46791$ \\
	      -0.211513 & $-0.243935,  0.455242 \pm 1.65563 i, 2.60442$ \\
	        0.182165 & $-0.257897, 0.455108 \pm 1.65296 i, 2.60409 $\\
		 1.02751 & $-0.5 + 1.49328 i, -0.247673, 0.810822, 2.66003$ \\
		   1.17576 & $-0.5 + 1.48929 i, -0.252564, 0.805474, 2.65973$\\
		    2.29919 & $-0.249424, 0.754748, 2.26241 \pm  0.572603 i$ \\
		     2.33595 & $-0.250589, 0.753578, 2.26201 \pm  0.571944 i$\\
		      5.66788  & $-0.5 + 0.392135 i, 0.451997  \pm 1.56424 i, 2.59575 $\\
	  8.13696  & $0.000191345 \pm 0.337831 i, 1.62674, 2.36175 $\\
	  9.45442  & $-0.5 + 1.00956 i, 0.694733 \pm 1.28851 i, 2.49691 $\\
    10.8455  & $0.613599 \pm 3.18391 i, 2.74696 \pm 2.01537 i$ \\ 
    \hline
   \end{tabular}
   \caption{The complete set of $2^{N}$ energy levels and Bethe roots for 
   $N=4$
   using Eqs.  (\ref{TQ2})-(\ref{Q}) and boundary parameter values $p=1/4\,, q=1/2\,, \xi=-\sqrt{3}$. }
   \label{table:N=4}
\end{table}

\end{document}